SURFACE, THIN FILMS

# FEASIBILITY OF STUDY MAGNETIC PROXIMITY EFFECTS IN BILAYER "SUPERCONDUCTOR/FERROMAGNET" USING WAVEGUIDE-ENHANCED POLARIZED NEUTRON REFLECTOMETRY

Yu.N. Khaydukov[a], Yu.V. Nikitenko[a], L. Bottyan[c], A. Rühm[d], V.L. Aksenov[a]

[a] *Joint Institute for Nuclear Research, Joliot Curie 6, Dubna, 141980 Russia*

[c] *Research institute for particle and nuclear physics, Budapest, H-1121 Hungary*

[d] *Max-Planck-Institut für Metallforschung, Heisenberg 3, Stuttgart, D 70569 Germany*

E-mail: khaiduk@nf.jinr.ru

Abstract

A resonant enhancement of the neutron standing waves is proposed to use in order to increase the magnetic neutron scattering from a "superconductor/ferromagnet"(S/F) bilayer. The model calculations show that usage of this effect allows to increase the magnetic scattering intensity by factor of hundreds. Aspects related to the growth procedure (order of deposition, roughness of the layers etc) as well as experimental conditions (resolution, polarization of the neutron beam, background etc) are also discussed.

Collected experimental data for the S/F heterostructure Cu(32nm)/V(40nm)/Fe(1nm)/MgO confirmed the presence of a resonant 60-fold amplification of the magnetic scattering.

## 1. INTRODUCTION

In the physics of ferromagnet (F)/superconductor (S) proximity effects there are a number of intriguing predictions requiring further clarification and experimental verification [1]. In particular, it was predicted that due to the proximity effect various scenarios of the influence of the superconducting phase transition on the magnetic properties of the layered F/S systems are possible: the formation of inhomogeneously ordered magnetic structure (cryptoferromagnetism)[2], magnetization "leakage" from F to S layer (inverse proximity effect)[ 3-5] etc.

For the experimental study of the problem various methods such as SQUID magnetometry [6], resonant synchrotron radiation scattering [7], ferromagnetic resonance [6,8] and some others are extensively used. For example, ferromagnetic resonance was used in [6,8] for studying the behavior of the saturation magnetization of the F layer near the superconducting transition point. Decreasing of the saturation magnetization value (the effect is from 4 up to 60%) has been found at superconducting transition. This decreasing can be explained either by transition into a domain state or by the inverse proximity effects.

A promising method for the investigation of this problem is Polarized Neutron Reflectometry (PNR). The advantage of this method is the ability to obtain detailed information about the spatial distribution of the magnetic induction in the structure. In this case the achievable spatial resolution is of the order of 1 nm and less. In the works [9,10] PNR method was used for the investigation of multilayered (n>3) S/F heterostructures. However, because of the big number of S/F interfaces, the interpretation of the obtained data was difficult. Studying of more simple (two-and three-layer) system would simplify the task. However, in this case the intensity of the signal is rather low due to small amount of magnetic material. For example, in [11] we used PNR to study the three-layer S/F/S system. The anomalously high suppression of the mean magnetization in the F layer (effect is around 90%) was found below superconducting transition temperature $T_C$. Such a decrease can be attributed to the decrease of saturation magnetization or to the change of the domain state in the structure. The definitive answer could be obtained after a quantitative analysis of the neutron diffuse scattering, which in that work has not been made because of its low intensity.

In the present work we propose to use a resonant enhancement of the neutron standing waves [12] for a significant increasing of the magnetic scattering intensity from a simple S/F structure. Performed model calculations were done for the S/F bilayer, but can also be applied for S/F/S and the F/S/F three-layered heterostructures. Experimental investigation has been carried out on Cu/V/Fe/MgO system, where the investigated S/F bilayer iron/vanadium is placed between Cu and MgO layers creating the resonant enhancement.

## 2. SCHEME OF THE PNR EXPERIMENT

The scheme of the typical PNR experiment is shown on Fig.1. Neutron beam with wavelength $\lambda$ is polarized by the polarizer P. The principle of operation of the polarizer is based on the dependence of the transmission coefficient of neutron on the spin direction. As a polarizer magnetic supermirrors [13] or polarized $^3He$ gas can be used [14,15]. Further, the polarized beam is collimated by the system of slits and directed to the sample SM under the angle of incidence $\theta_1 \sim 10$ mrad. The outgoing polarization of the scattered neutron beam under the angle $\theta_2$ is analyzed by an analyzer of polarization AP. The principle of operation is the same as for polarizer. After that the scattered intensity is recorded by the detector D. The alternation of the beam polarization direction to the guide magnetic field before and after the sample is realized by the spin-flippers $SF_1$ and $SF_2$. Spin-flipper is a special device consiting of the combination of electromagnetic coils, creating a certain configuration of magnetic field. When a spin-flipper is switched on, the direction of neutron beam polarization reverses.

The measured quantities in PNR experiments are intensities of specular reflection ($\theta_2 = \theta_1$) and diffuse scattering ($\theta_2 \neq \theta_1$) at different angles of incidence and states of the spin-flippers (on-on, off-off, on-off and off-on).

The primary data treatment procedure consists of taking into account the efficiencies of the polarizer $P_p$, analyzer $P_a$, spin flippers $P_{sf1\ (sf2)}$ and background intensity $I_b$ [16]. The result of this processing is the reflectivities $R^{\mu\eta}(Q)$ as a function of momentum transfer Q, normalized on the intensity of the incident beam. Here indices $\mu$ and $\eta$ take values "+" or "-" and correspond to the projection of the neutron spin on the guide field before and after scattering process respectively. Experimental reflectivities are compared with the theoretical values calculated for the model potential of interaction for the investigated structure. The interaction potential depends on nuclear and magnetic contributions and for the case Q<<10 $nm^{-1}$ can be written in the approximation of continuous medium [17]:

$V(\vec{r}) = 4\pi\rho(\vec{r}) + C\boldsymbol{\sigma}\mathbf{B}(\vec{r})$, where (1)

$\rho$ - nuclear potential of the interaction (or nuclear scattering length density, SLD) and **B(r)** – distribution of the magnetic induction in the structure, $C = -\frac{2m}{\hbar^2}\mu_n = 2.91\cdot10^{-6}\,kOe\cdot A^{-2}$, m – neutron mass, $\mu_n$ – neutron magnetic moment. In order to fit experimental data one should take into account the following points. Specular reflection allows reconstruct depth-profile of the SLD. If nuclear and/or magnetic in-plane inhomogeneties are present in the structure (roughness, clusters, domains etc), they will cause diffuse scattering. Depth profile in this case is obtained by averaging in the plane on the size equal to the coherence length of neutron beam $L_{coh}$. The latter is inversely proportional to the uncertainty of the momentum transfer dQ and varies in the range from 0.1 micron to 1 centimeter [18]. Specular reflectivities $R^{++}$ and $R^{--}$ depend on the nuclear SLD contribution and component of the magnetic induction $B_{||} = |B|\cdot\cos\alpha$ parallel to the external magnetic field:

$$R^{\overset{++}{(--)}} \sim \left|4\pi\rho_0(z) \pm C B_{||}(z)\right|^2 \qquad (2)$$

Spin-flip reflectivities depend on perpendicular component of the magnetic induction $B_\perp = |B|\cdot\sin\alpha$ i.e. it is solely of magnetic origin:

$$R^{\overset{+-}{(-+)}} \sim |\ C B_\perp(z)|^2 \qquad (3)$$

Here and later α - is the angle between the external magnetic field direction and magnetic induction in the structure. For a weak-magnet system ($|CB_{||}| << |4\pi\rho|$) reflectivities $R^{++}$ and $R^{--}$ are mainly sensitive to the nuclear part of the potential. In order to substract profile $B_{||}(z)$, it is convenient to use the so-called spin asymmetry $S \equiv [R^{++} - R^{--}]/[R^{++} + R^{--}]$, as it can be shown from (2) that $S \sim B_{||}(z)$.

## 3. THEORY OF RESONANT ENHANCEMENT OF MAGNETIC SCATTERING

The intensity of spin-flip scattering of a thin magnetic layer with thickness Δd can be written as [12]

$$I^{\overset{+-}{(-+)}}(k_{fz}, k_{iz}) \propto \psi^+(k_{fz})\psi^-(k_{iz})\Delta d^2 B_\perp^2 \qquad (4)$$

Here $k_{iz}$ ($k_{fz}$) = $\pm 2\pi \sin(\theta_{i(f)})/\lambda$ is the z-component of the incident (scattered) wavevector, $\psi^{+(-)}$ - neutron wave function with given polarization in the position of magnetic layer. The square of the wave function determines the density of neutrons with a given polarization near the thin layer. For the specular reflection we have $k_{fz} = - k_{iz}$, $Q_Z = 2k_{iz}$, $Q_{||} = 0$. A similar expression can be written for the intensity of diffuse scattering from magnetic inhomogeneities in the layer [19]:

$$I_d(k_{fz}, k_{iz}, Q_{||}) \propto \psi(k_{fz})\psi(k_{iz})\Delta d^2 \int d^2 r_{||} \exp(-Q_{||} r_{||}) \langle B(r_{||}), B(0) \rangle, \qquad (5)$$

where $Q_{||}$, $r_{||}$ - components of the momentum transfer and radius vector in-plane of the structure, sign «<>» means correlator.

From the expressions (4) and (5) it follows that intensity of the magnetic scattering depends on neutron density close to magnetic layer. If we put it in the resonator, under the certain conditions, the density of neutrons and, consequently, the scattering intensity will be increased substantially. For the system depicted on Fig. 2 the wave function near the magnetic layer can be written as [20]

$$\psi^{\pm} = \frac{Te^{ik_2 d_2}}{1 - r_{21} r_{23}^{\pm} e^{2ik_2 d_2}}, \text{ where} \qquad (6)$$

T – transmission amplitude from the 1st layer, $r_{21}$ and $r^{\pm}_{23}$ are reflection amplitudes from the interfaces 1/2 and 2/3, $k_i = \sqrt{k_0^2 - V_i}$ is the z-component of the wavevector inside the i-th layer(i=1,2 or 3). From the expression (6) follows that wave function is resonantly increased when the following conditions fulfilled:

$$\arg(r_{21}) + \arg(r^{\pm}_{23}) + 2\mathrm{Re}\,(k_2 d_2) = 2\pi n, \qquad (7.1)$$

$$|r_{12}| = |r_{23}| = 1. \qquad (7.2)$$

A characteristic property of the resonator with non-collinear magnetic layer is that the amplitude of reflection $r^{\pm}_{23}$ depends on the magnitude of the magnetic induction in the structure and is different for the "+" and "-" spin-state [20]. As a sequence conditions (7.1) for $\psi^+$ and $\psi^-$ functions will be observed at different wavevectors. This leads to the fact that the resonance in the spin-flip channel consists of two adjacent peaks. The distance between the peaks can be approximately written as [20]

$$\Delta Q \approx \frac{4(CB)^2 \Delta d\, d_2^2}{\pi^3} \qquad (8)$$

The aim of this work is the application of the regime of resonant enhancement of neutron standing waves for the amplification of intensity of magnetic scattering from S/F bilayer. In order to reach this aim the following requirements should be fulfilled:

i) high optical contrast at interfaces 1/2 and 2/3 (see Fig.2(insert))

ii) S/F bilayer consisting of homogeneous and thin (~ 1nm) ferromagnet and thick (~ 100 nm) superconductor

iii) The lattice parameters and the space group for the ferromagnetic and superconducting materials should be similar, as well as resonator structure.

Condition i) is needed to provide total reflection at the interfaces 2/1 and 2/3 (see expression 7.2); term ii) is necessary for proximity effects existence itself, iii) is required for high-quality heterostructures with sharp interfaces to be prepared. There are several materials exist matching the conditions given above such as Pb,

Nb or V as a superconducting and Fe, Ni or Co as a ferromagnetic layer [21]. In the present work system $Cu(d_1)/V(d_2)/Fe(\sim 1nm)/MgO$ have been chosen. Here substrate, magnesium oxide ($\rho_3 = 6.0\times10^{-6}$ Å$^{-2}$), is widely used for producing epitaxial iron-vanadium heterostructures [22]. Mismatch of the lattice parameters for Fe and V crystals does not exceed 5%, allowing create high-quality heterostructures with ultra-low roughness (1-2 monolayers). The sputtering of iron directly on the substrate provides a thin, yet uniform ferromagnetic layer. Vanadium layer in this structure combines the properties of the superconductor and the phase-shifting layer as it has negative scattering length density ($\rho_2 = -0.27\times10^{-6}$ Å$^{-2}$). The sample-design procedure must take into account the dependence of the superconducting transition temperature of vanadium layer on its thickness $T_C(d_2)$ [23]. At $d_2 < 20$ nm $T_C = 0$, at $d_2 > 100$ nm $T_C$ of vanadium layer practically does not differ from the temperature of the bulk vanadium $T_{C,bulk} = 5.2$K. As a first reflective layer copper with nuclear SLD $\rho_1 = 6.5 \times 10^{-6}$ Å$^{-2}$ is used.

In order to get maximum enhancement factor for neutron magnetic scattering a numerical optimization of the thicknesses $d_1$ and $d_2$ has been done. For calculations we have chosen the F layer thickness $\Delta d = 1$ nm with a magnetic induction B = 10 kGs directed perpendicularly to the external magnetic field ($\alpha = 90^o$). The dependence of maximum spin-flip signal $R^{+-}_{max}$ on the thickness $d_1$ of copper layer and vanadium layer $d_2$ is depicted on Fig. 3-a. From this dependence the thicknesses were chosen as follows: $d_1(Cu) = 32$ nm and $d_2(V) = 40$ nm. Specular reflectivities for the optimized structure Cu(32nm)/V(40nm)/Fe(1nm)/MgO are shown in Fig. 3-b. The non-spin-flip reflectivities $R^{++}$ and $R^{--}$ are described by the total reflection plateau at $Q<QB_{cr} = 0.17$ nm$^{-1}$ and oscillations at $Q>Q_{cr}$ due to interference on thicknesses of Cu and V. Main features of the spin flip reflectivities $R^{+-}$ and $R^{-+}$ are the resonant peaks at $Q_{res} = 0.09$ nm$^{-1}$ (value of signal 70%) and close to $Q_{cr}$ (value of signal 50%). Presence of the resonance enhancement at $Q = Q_{res}$ make it possible to increase spin-flip signal 160 times (compare curves (2) and (3) in Fig. 3-b).

From the inset to Fig. 3-b it is seen that resonance for non-collinear magnetic system consists of two adjacent peaks. As it can be seen from (8), the distance between the peaks depends on the absolute value of magnetic induction in the S/F bilayer. On Fig. 4-a spin-flip reflectivities $R^{+-}$ near the resonance for the F layer with different absolute values of induction at $\alpha = 90^{o}$ are shown. It can be seen that an increase of the absolute magnitude of induction causes increase in distance between the peaks. Similar picture is observed if one adds additional magnetization in the S layer. Thus, the distance between the resonance peaks is a measure of integrated magnetic induction in the S/F bilayer. One should mention that a simple rotation of the magnetic induction does not change the distance between the peaks but only the intensity of the peaks (Fig. 4-b). We note here that similar effects can be also observed in the diffuse scattering channel therefore we won't go into details on it. Thus, analysis of the magnetic scattering intensity in the resonance peaks gives information about evolution of the magnetic state in the system. Apart from the conventional reflectometry, the curves are measured in a narrow Q range, near the resonance. This allows in short term (several hours) record information about the magnetic state of the system caused by the superconducting transition with high statistical accuracy.

## 4. INFLUENCE OF EXPERIMENTAL PARAMETERS ON THE OBSERVED EFFECTS

In the problem considered above we assumed scattering of the ideally polarized monochromatic neutron beam from the structure with the sharp interfaces. Let us consider now, how the real procedure of sample preparation and experimental conditions affect on observation of the resonances.

### 4.1. Roughness at the interfaces

At preparation of real structures presence of roughness at interfaces is inevitable. At specular reflection roughness can be taken into account using Debye-Waller or Nevot-Croce [24] factors. For reflection and transmission amplitude at i-1/i interface the Nevot-Croce factors can be written as $f_r=\exp(-2k_{i-1}k_i\sigma_i)$ and $f_t=\exp[(k_i-k_{i-1})^2\sigma_i^2/2]$ respectively. Here $\sigma_i$ – is the root-mean-square height of roughness on i-1/i interface. Far from the total reflection region parameter $f_r$ describes suppression of reflectivity due to presence of roughness. However, as the resonance take place below $Q_{cr,}$ values $k_1$ and $k_3$ are imaginary and, therefore factor $f_r$ is also imaginary for interfaces 2/1 (V/Cu) and 2/3 (V/[Fe/MgO]). This means that according to (7.1) parameters $\sigma_2$ and $\sigma_3$ can only shift position of the resonant peak. The only parameter that can suppress resonance is the roughness at surface, as transmission T depends on it. Direct calculations show that condition $\sigma_1 \ll 1/Q_{res}=10$nm is needed to save resonant enhancement.

## 4.2 Instrumental resolution

Intensity measured in the experiment is the convolution of the true intensity with the resolution function of the instrument. The latter can be represented by a Gaussian with width ΔQ, associated with the uncertainty of momentum transfer, which depends on uncertainty in determining of the wavelength Δλ and angular divergence Δθ:

$$(\Delta Q/Q)^2 = (\Delta\lambda/\lambda)^2 + (\Delta\theta/\theta)^2 . \qquad (9)$$

Uncertainty Δλ is connected with the properties of the monochromator crystal for monochromatic mode or with a time resolution in the case of time-of-flight reflectometers. Usually this value is small and does not exceed 1%. The angular divergence of the beam is given by the size of collimating slits and may vary over a wide range from 1% to 30%. Therefore, resolution of a neutron reflectometer is of the order ΔQ/Q ~ 1-10%, which does not allow resolving the resonance peaks separately. Each resonance will be represented by one peak only. Relation of the parameters of this peak with the magnetic state of the S/F bilayer is shown in Table 1.

The experimental resolution should be chosen based on the following considerations: decreasing of ΔQ increases sensitivity to the change of peak parameters and therefore to the changes in magnetic state of S/F bilayer. However decreasing of $\Delta Q$ is reached by reducing the slit sizes, thereby decreasing the intensity of the incident beam. On the other hand, increasing in $\Delta Q$ would smear peak out and, starting from a certain value, peak becomes smaller than the background intensity. For the system being studied presently, the optimum resolution is found to be $\Delta Q/Q_{res} = [2 \div 6]$ %.

4.3 Background and finite efficiency of the polarizing/analyzing elements

Polarization degree of the neutron beam also affects on observation of resonances. Intensity of the measured spin-flip peak for the case $P_p < 1$ and $P_a = P_{sf1} = P_{sf2} = 1$ can be written as [16]

$$I^{off\text{-}on(on\text{-}off)} = \tfrac{1}{2}\left[R^{+-}(1 + P_p) + R^{++}(1 - P_p)\right], \tag{10}$$

from which follows that term $R^{++}(1 - P_p)$ works as background. Signal-to-noise ratio makes value

$$signal/noise = R^{+-}/R^{++} \times (1 + P_p)/(1 - P_p) \tag{11}$$

To have the condition signal/noise>1 one must require $P_p > 1 - 2R^{+-}$ [26]. For example, if one observes the resonance peaks of about 10% (related to the total reflection intensity) it is required that the polarization of the beam must be around 80%. If the other values like $P_a$, $P_{sf1}$ or $P_{sf2}$ are significantly less than 1, then polarization must be even higher.

## 5. EXPERIMENTAL DATA

Sample with nominal composition Cu(32nm)/V(40nm)/Fe(1nm)/MgO(001) (size ~20×10×2 mm$^3$) have been prepared in the KFKI Research Institute for Particle and Nuclear Physics (Budapest, Hungary) by molecular beam epitaxy at a temperature of the substrate 300° C in vacuum of $2.2\times10^{-10}$ mbar [26]. Structural properties and layers composition have been investigated by X-ray diffraction (both large and small angle), Rutherford back-scattering (RBS) and Secondary Neutral Mass Spectrometry. Measurements have shown high-quality of the layers and interfaces: rms-height of roughness on MgO/Fe and Fe/V interfaces are estimated to be less than 0.6 nm. In the case of Cu/V interface, roughness was higher (around 2-5 nm) due to difference in cell parameters. Magnetic properties of the F layer at low temperatures were investigated by SQUID magnetometry at the KFKI institute of Solid Physics and Optics (Budapest, Hungary). Measurements have proven ferromagnetic state of the iron layer with coercivity and saturation field $H_{cr}$ = 35 Oe and $H_{sat}$ =0.5 kOe, respectively. The saturation magnetization of the F layer is $B_{sat}$ = 17.5 kGs which is 80% of the bulk value of the iron. At H=0 magnetization suppressed by 10% in comparison to $B_{sat}$. Such a small decrease points out on that magnetic state of the F layer at remanence is close to the homogeneous state in the saturation.

The first series of experiments were carried out on monochromatic ($\lambda$ = 5.5 Å) reflectometer NREX + at reactor FRM-II (Munich, Germany). The polarization of the beam was $P_p$ = 93%. A cell with polarized $^3$He gas with initial efficiency $P_a$ = 75-80% was used as an analyzer for the scattered neutron beam polarization. Reflectivity curves $R^{++}$, $R^{--}$ and $R^{+-}$ have been measured at room temperature in the range of momentum transfer values Q = 0.07÷0.3 nm$^{-1}$.

The resonant enhancement was checked by the presence of peak in the spin-flip channel. As spin-flip depends on non-collinear component of the magnetization (see expressions 3 and 4) following protocol was used to increase it:

a) applying external field $H > H_{sat}$ for 1 minute and release it, in order to make homogeneous magnetic state,

b) rotating the sample by 90°,

c) applying of small guiding magnetic field to keep neutrons polarization.

If the guide field is less than coercivity of F layer, magnetization remains directed non-collinearly to the external field. Using this procedure, at external field $H \approx 20$ Oe, we were able to observe the resonance at $Q_{res}$ = 0.09 nm$^{-1}$ with an intensity of about 10% relative to the total reflection intensity (fig. 5-a).

Data fit has been done in several stages. Firstly, the reflectivity curves $R^{++}$ and $R^{--}$ have been fitted by varying thicknesses of the layers and its nuclear SLD's only. This provided the initial distribution of the SLD. On the second stage spin asymmetry and spin flip $R^{+-}$ reflectivity were fitted by varying the magnetic parameters of the F layer such as absolute value of magnetization and the angle between magnetic induction in the layer and the guide field. Simultaneous fit allowed to obtain absolute value of magnetization in the F layer at room temperature $|B| = 11 \pm 1$ кGs and the angle $\alpha = 35^{o} \pm 5^{o}$. During the fit procedure on this stage, the nuclear profile was also corrected as the position and the absolute value of the resonant spin-flip peak depends on thicknesses and SLDs of vanadium and copper layers. Finally obtained nuclear profile allowed to calculate neutron density in the structure at resonance (Fig. 5-b). According to the calculations, neutron density near the FM layer increases by factor n = 50 compared to the incident flux. Such enhancement allowed to increase of spin-flip signal by factor 60 for the given instrumental resolution.

## 6. CONCLUSIONS

For tracing the evolution of the magnetic state in the Cu(32nm)/V(40nm)/Fe(1nm)/MgO(001) heterostructure of "superconductor/ferromagnet" type it is proposed to use a resonant enhancement of neutron standing waves. Our calculations show that the usage of this effect allows to increase the intensity of magnetic scattering of polarized neutrons by several hundreds times. This gives an opportunity to measure weak changes of the magnetic state of S/F bilayer due to proximity effects. The requirements to the technology of preparation (the order of deposition, the roughness of layers etc) and the technique of the experiment (resolution, the degree of polarization of the beam, background intensity) are discussed.

Experimental data for the structure Cu(32nm)/V(40nm)/Fe(1nm)/MgO confirmed the resonant amplification by factor 60 of the neutron magnetic scattering at $Q_{res} = 0.09$nm.

ACKNOWLEDGEMENTS

The authors would like to thank L. Deak, F. Tanczikó, K. Gábor, D. Merkel and B. Nagy (KFKI Research Institute for Particle and Nuclear Physics) for the fruitful discussions, samples preparation and assistance in PNR measurements at NREX+,K. Zhernenkov (FLNP JINR) for the fruitful discussions. The work was done under the financial support of the Russian Foundation for Basic Research (grant № 09-02-00566) and JINR-HAS (EAI-2009/002). Yu. Khaydukov would like to acknowledge the support of the Foundation for Assistance to Small Innovative Enterprises of the Russian Federation (grant № 8455 UMNIK-08-3).

TABLES

Table 1. Relation of the parameters of experimentally observed resonant spin-flip peak with magnetic state of S/F bilayer.

| Parameter of the peak | Parameter of the structure |
|---|---|
| Peak width (W) ( = distance between non-resolved peaks) | Absolute value of induction \|B\| in S/F bilayer |
| Peak height (H) | perpendicular component $B_{\perp}$ and absolute value \|B\| in S/F bilayer |
| Peak position ($X_C$) | nuclear potential |

FIGURE CAPTIONS

Fig 1. PNR experiment scheme. P – polarizer, SL – slit, $SF_{1(2)}$ – spin-flippers, H – external magnetic field, SM – sample, AP – analyzer of the polarization, D – detector

Fig. 2. Scattering of polarized neutrons from resonant structure with the thin non-collinear layer. On figure $k_{i(f)}$ – incident (scattered) wavevector, $\theta_{i(f)}$,-incident(final) angle, $Q_{z(||)}$ - perpendicular (parallel) component of the momentum transfer Q, **B** – magnetic induction in the FM layer, $\alpha$ - angle between induction and external magnetic field, Nuclear SLD profile of the structure is shown in the right side of the figure.

Fig. 3. a) The dependence of the maximum spin-flip signal from the thickness $d_1$ of copper and vanadium $d_2$. b) - $R^{++}(Q)$ (1) and $R^{+-}(Q)$ (2) reflectivities for an optimized system Cu(32nm)/V(40nm)/Fe(1nm)/MgO. Inset - curve (2) near the resonance. (3) - The dependence of $R^{+-}(Q)$ for V(40nm)/Fe(1nm) bilayer without a resonator.

Fig. 4. The dependence of the resonant spin-flip signal from the absolute value (a) and direction (b) of magnetic induction in the F layer.

Fig. 5. a) – The hysteresis loop of the FM layer measured by SQUID at T=10K. b) – Value of upper critical field $H_{C2}$ of superconductivity applied in-plane of the sample as a function of the temperature (dots). Solid line – fit using expression (9) with parameters $T_C$ = 3.4K, $d_2$ = 35 nm, $\xi(0)$ = 15 nm.

Fig. 6. The results of room temperature PNR experiment. a) Experimental (points) and model (solid lines) curves of specular reflection in the different channels. Position of the resonance is shown by arrow. Inset - the spin asymmetry. b) Nuclear profile reconstructed after data fitting and neutron density in the resonance corresponding to this profile.

INFORMATION ABOUT THE AUTHORS

1) Yury N.Khaydukov – correspondence author

141980 Dubna, Russia, FLNP JINR

Tel. +7-49621-62875 - office

Fax: +7-49621-65484,

e-mail: khaiduk@nf.jinr.ru

3) Yury V. Nikitenko

141980 Dubna, Russia, FLNP JINR

Tel. +7-49621-65155 - office

Fax: +7-49621-65484,

e-mail: nikiten@nf.jinr.ru

5) Laszlo Bottyan,

H-1121, Budapest, Hungary, Research institute for particle and nuclear physics.

Tel. +361 3922761 – office,

Fax: +361 3959151,

e-mail: bottyan@rmki.kfki.hu

6) Adrian Rühm

D 70569, D-85747 Garching bei München, Germany

Tel: (+49) (0) 89 289 – 14878 – office

Fax: (+49) (0) 89 289 – 14989

E-mail: ruehm@mf.mpg.de

9) Victor L. Aksenov

141980 Dubna, Russia, FLNP JINR

Tel. +7-49621-66844 - office

Fax: +7-49621-65484,

e-mail: aksenov@nf.jinr.ru

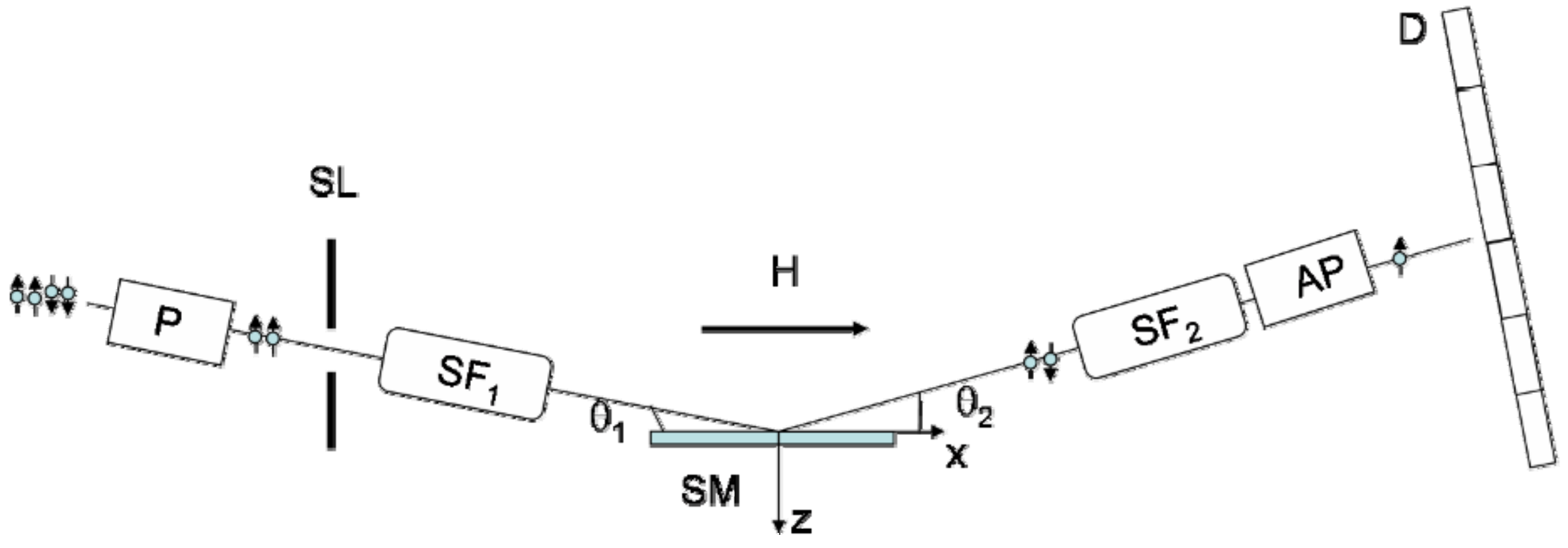

Fig. 1

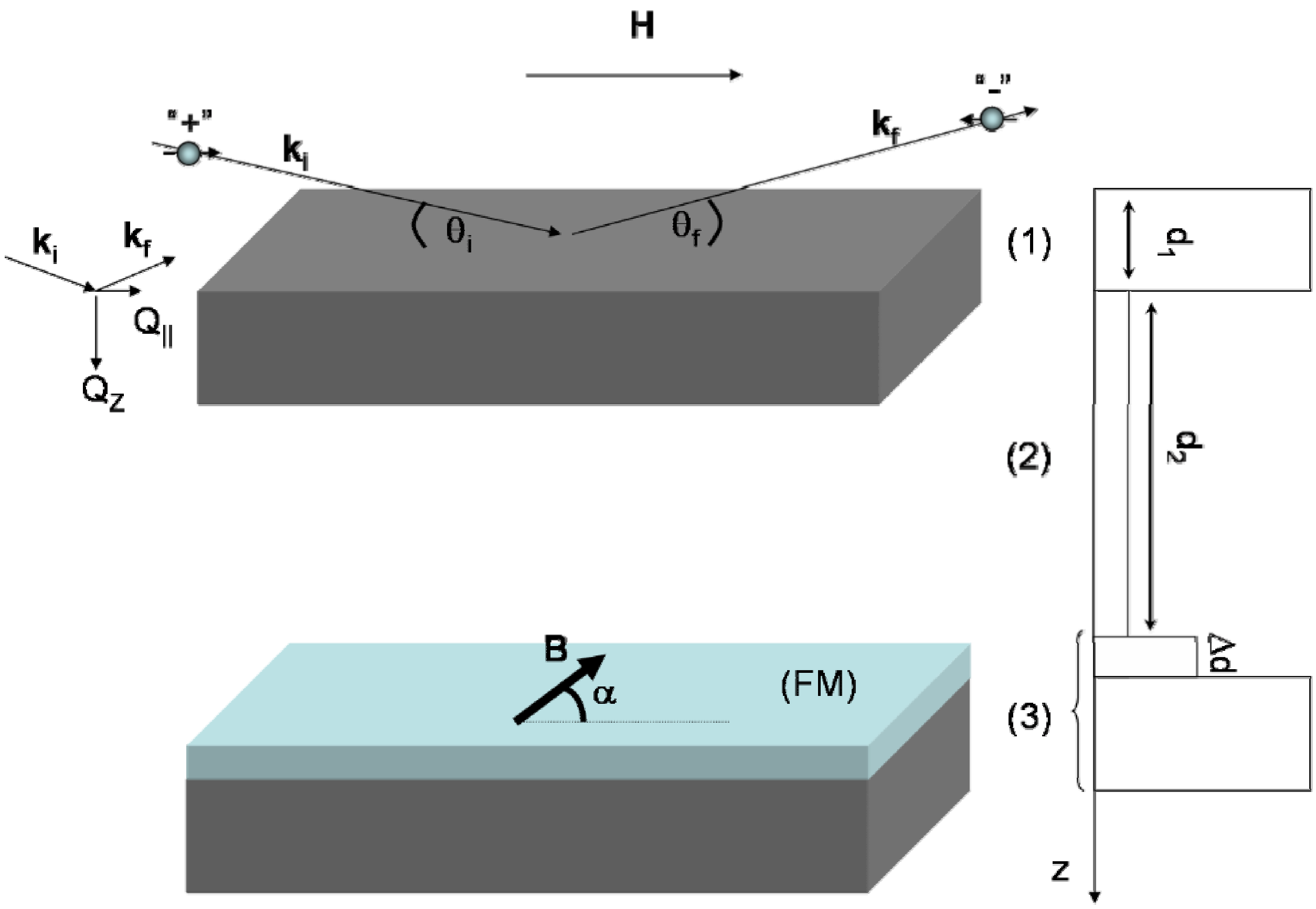

Fig. 2

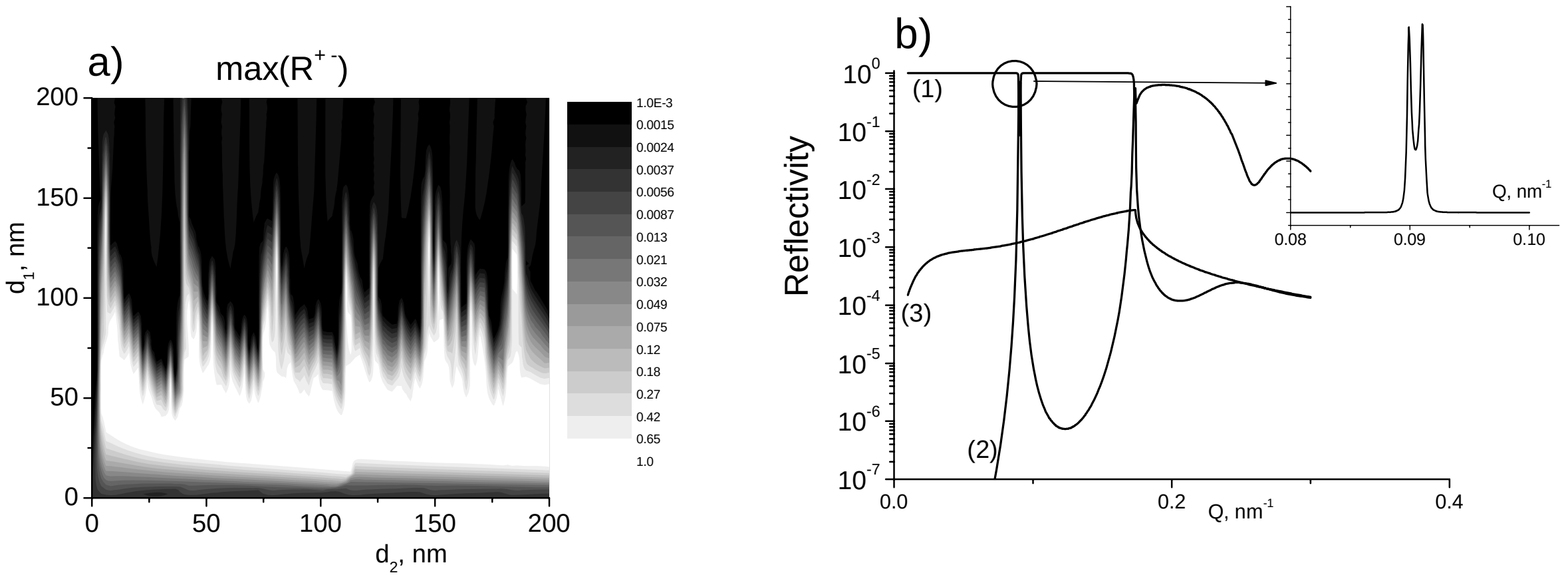

Fig. 3

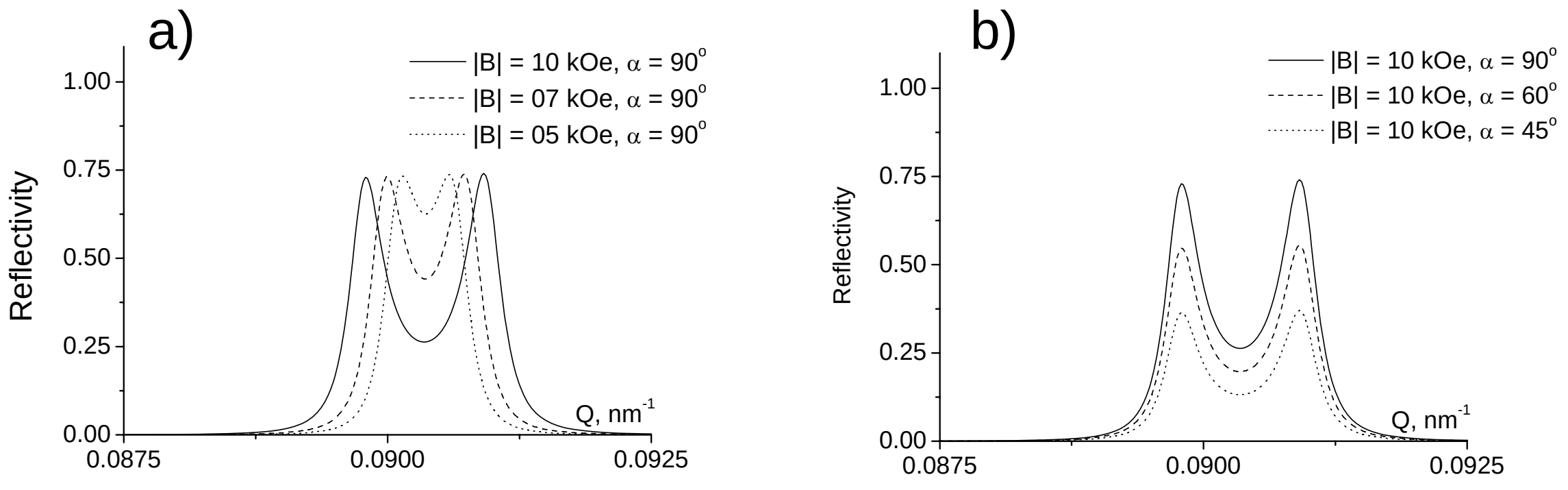

a)
|B| = 10 kOe, α = 90°
|B| = 07 kOe, α = 90°
|B| = 05 kOe, α = 90°
Reflectivity
1.00
0.75
0.50
0.25
0.00
Q, nm-1
0.0875
0.0900
0.0925
b)
|B| = 10 kOe, α = 90°
|B| = 10 kOe, α = 60°
|B| = 10 kOe, α = 45°
Reflectivity
1.00
0.75
0.50
0.25
0.00
Q, nm-1
0.0875
0.0900
0.0925

Fig. 4

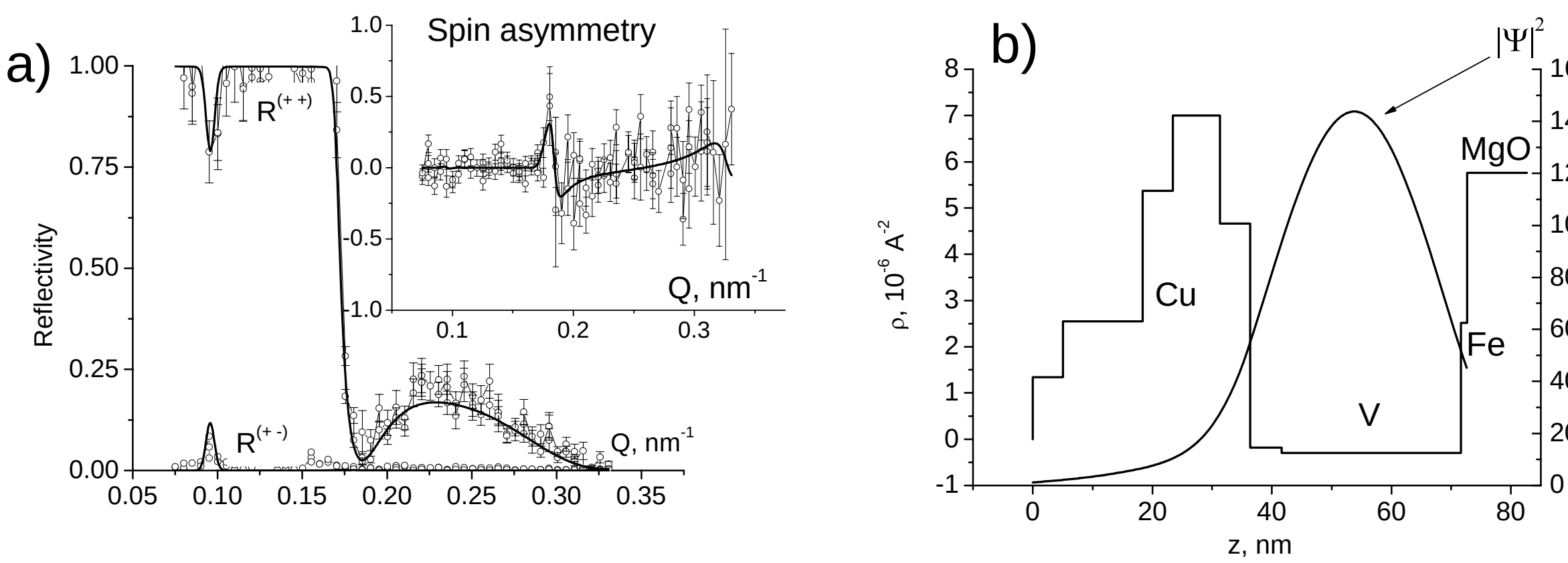

Fig. 5